\documentclass[aps,prb,showpacs,byrevtex,showpacs,amsmath,amssymb,twocolumn]{revtex4}
\usepackage{amssymb,dsfont}

\begin{document}
\title{The Hubbard model with spin orbit coupling: \\a lattice gauge theory approach}
\author{Giuseppe Guarnaccia}
\affiliation{Dipartimento di Fisica ``E.~R. Caianiello'', Universit\`a di
Salerno, I-84084 Fisciano (Salerno), Italy}

\author{Canio Noce}
\affiliation{SPIN-CNR, I-84084 Fisciano (Salerno), Italy}
\affiliation{Dipartimento di Fisica ``E.~R. Caianiello'', Universit\`a di
Salerno, I-84084 Fisciano (Salerno), Italy}

\begin{abstract}

We study the symmetry properties of the Hubbard model with spin-orbit interactions of Rashba and Dresselhaus type. These interactions break the rotational symmetry in spin space, so that the magnetic order cannot be excluded by using the Bogoliubov inequality method. Nevertheless, we rigorously show that the existence of the magnetic long-range orders may be ruled out when the Rashba and Dresselhaus coupling constants are equal in modulus, whereas the $\eta$-pairing can be always ruled out, regardless of the microscopic parameters of the model. These results are obtained by imposing locally the SU(2) gauge symmetry on the lattice, and rewriting the spin-orbit interactions in such a way that they are included in the path ordered of the gauge field on lattice.

\end{abstract}
\maketitle

\section{Introduction}
One of the most interesting features of the interacting systems is the existence of a macroscopic order which may break the underlying symmetry of the Hamiltonian. For instance, it is known that the continuous rotational symmetry in the three-dimensional spin space of the isotropic Heisenberg ferromagnet is broken by the spontaneous magnetization that exists in the limit of vanishing magnetic field for a three-dimensional lattice.\cite{Dys} For systems of restricted dimensionality and at any finite temperature, it has been argued that there is no macroscopic order, both on the basis of heuristic arguments and by a rigorous approach based on the use of an operator inequality due to Bogoliubov.~\cite{bog} The piece of work by Bogoliubov stimulated numerous investigations on the upper and lower bounds for thermodynamic averages, stating that spontaneous ordering is usually not possible in low-dimensional systems.~\cite{Nolt}

When a lattice chain of spins is considered, pioneering work by Mermin and Wagner~\cite{MW} ruled out magnetic ordering both in one-dimensional (1D) and two-dimensional (2D) quantum spin systems with an isotropic (Heisenberg) interaction between the spins, at any nonzero temperature. There are, however, other interactions, such as the Ruderman-Kittel-Kasuya-Yosida interaction (RKKY),~\cite{RKKY} in which the conduction electrons mediate the interaction between spins. These interactions are long-ranged and hence not covered by the Mermin-Wagner result. Theoretical work has shown that ordering is not possible in 1D RKKY systems,~\cite{Bruno} but nothing similar has yet been attempted for the 2D case. Nevertheless, recently it has been shown that the Mermin-Wagner theorem can be applied also to systems of lattice spins which are spin coupled to itinerant and interacting charge carriers.~\cite{Fes} Within this approach,  it has been rigorously proved that neither (anti-)ferromagnetic nor helical long-range order is possible in 1D and 2D, at any finite temperature. This conclusion is not related to a specific model Hamiltonian but applies to a wide class of models including any form of spin-independent electron-electron and single-electron interactions.

Nevertheless, this argument becomes inconclusive when interactions which explicitly break the spin symmetry are considered. In this case magnetic order cannot be excluded. This is the case for instance of spin-orbit interactions (SOIs) such as the Rashba~\cite{Ras} or the Dresselhaus~\cite{Dress} ones. We recall that the Rashba spin-orbit field can be attributed to an electric field that originates from a structural inversion asymmetry whereas the Dresselhaus interaction comes from bulk inversion asymmetry.~\cite{Mai}
Since both these spin-orbit interactions break the symmetry of the spin interaction, their presence means that the magnetic order cannot be excluded. For completeness we mention that this conclusion is valid except for the special case in which the Rashba and Dresselhaus contributions are of equal magnitude.~\cite{Fes} This result turns out to be experimentally relevant because the Rashba spin-orbit interaction can be electrically tuned to match the Dresselhaus interaction, opening up possibilities for tuning magnetic order electrically.

Here, we investigate a specific model, namely the single-band Hubbard model, in presence of SOIs, and we rigorously show that the existence of long-range orders may be ruled out also when SOIs are present and the Bogoliubov inequality method is applied. Since SOIs break the spin symmetry of the model, we first introduce SOIs in the model by imposing local SU(2) gauge symmetry on the lattice, then we rewrite SOIs in such a way that they are included in the path ordered of the gauge field on the lattice. In this way, the  SU(2) invariance is restored allowing the use of the Bogoliubov inequality and the derivation of upper bounds for order parameters.
At this stage, it is worth mentioning that local symmetries, such as the local gauge symmetry,  cannot be broken spontaneously. Indeed, the Elitzur's theorem~\cite{El} states that a local gauge symmetry cannot be broken spontaneously, that is, the expectation value of any gauge non-invariant local observable (order parameter) must vanish. This result means that the spontaneous breaking of the gauge invariance can only occur when the local symmetry is explicitly broken by gauge fixing. Hence, first one chooses a gauge and in this gauge the remaining global gauge symmetry is spontaneously broken, as it happens for instance in the Anderson-Higgs mechanism.~\cite{Higgs} We also mention that for two-dimensional systems, if the residual symmetry is continuous, after the gauge fixing, at finite temperatures the associated order parameter must vanish according to the Mermin-Wagner theorem. Nevertheless, when a discrete symmetry is considered then the ordered phase can still exist. These results will be used in the next sections to describe the Hubbard model with SOIs within a SU(2) lattice gauge theory.

We point out that, recently the model here investigated has been extensively studied. Indeed, the  analysis of the properties of non-centrosymmetric superconductor CePt$_3$Si stems from the single-orbital Hubbard model with SOIs, since this model has been considered as a valid effective model for describing low-energy quasiparticles in the Fermi liquid state.~\cite{yanase04}. Moreover, experiments of the penetration depth of Li$_2$Pd$_3$B and Li$_2$Pt$_3$B reflect the difference of the strengths of the SOIs. Motivated by these data, the study of pairing symmetry of these non-centrosymmetric superconductors is again based on the single-band Hubbard model with SOIs.~\cite{yokoyama07}
New development of technology of epitaxial growth makes it possible to fabricate heterointerface between different transition-metal oxides. In SrTiO$_3$/LaAlO$_3$ the mobile electrons at the heterointerface are mainly introduced to 3d orbitals of Ti$^{3+}$ ions. Considering the crystalline electric field by a ionic model, the two-dimensional d$_{xy}$ orbital has the lowest orbital energy level in 3d manifold, as also shown by x-ray absorption spectroscopy with the linearly polarized light. To microscopically model these hybrid structures, a two-dimensional single-band Hubbard model with the SOIs as a minimal model of the two-dimensional electron gas at the heterointerface has been considered.~\cite{yada09}
Finally, we mention that a purely theoretical study of the interplay between electron-electron interactions has been recently performed again within a single-band Hubbard model with on-site, and eventually nearest-neighbor matrix elements of the Coulomb interaction in the presence of SOIs.~\cite{alexandradinata10}

The paper is organized as follows: in the next section we summarize some results on the lattice gauge theory, emphasizing their application to the Hubbard model presented in Sect. III. Section IV describes the application of the results of the previous sections and the main result of the paper, namely  the application of the Bogoliubov inequality to the Hubbard model in the presence of SOIs. The last section is finally devoted to remarks and conclusions.

\section{Review on Hamiltonian lattice gauge theory}
To straightforwardly understand the outcomes of the next section, here we summarize some important results of the lattice gauge theory.~\cite{KS} Since the interactions may be interpreted as a consequence of the local invariance of the theory under some gauge transformation groups, our aim is to show that the SOI may be included in the Hubbard model as the path ordered of gauge fields on the lattice.

If \textit{G} is a gauge transformation group, its elements are\\
$$g=e^{i\theta_aT_a}, $$\\
where $\theta_1,...,\theta_n$ are the group parameters and $T_1,...,T_n$ are the group generators. These operators fulfill the condition
$$[T_a,T_b]=if_{abc}T_c, $$
where $(f_{abc})_{a, b, c=1}^n$ are the group structure constants.

In quantum mechanics, the group transformations are realized by means of unitary operators $F$ in the Hilbert space $V_s$
$$\mid\alpha>\rightarrow\mid\alpha^\prime>=F\mid\alpha>, $$
for any $|\alpha>\in V_s$. Therefore, if $\Lambda$ and $\Psi(x)$ are the crystal lattice and a field operator, and $x\in\Lambda$, then\\
$$\Psi(x)\rightarrow F^+\Psi(x)F=e^{i\theta_aT_a}\Psi(x), $$\\
where the operator $F$ is\\
$$F=e^{i\theta_aQ_a}. $$\\
In this expression $Q_a=\sum_{x\in\Lambda}\Psi^+(x)T_a\Psi(x)$ with $a=1,...,n$ are the so-called charge operators and they are a representation of the group generators, thus implying that they verify the same commutation relation. If $[Q_a,H]=0$, then the quantum system has the global symmetry under the group considered.

A local transformation is defined by the following relation:\\
\begin{equation}
\label{a}
    F^+\Psi(x)F=e^{i\theta_a(x)T_a}\Psi(x).
\end{equation}\\

\noindent From these preliminary definitions, it is obvious that a system is globally gauge invariant if the model Hamiltonian contains  only terms like $\Psi^+(x)\Psi(y)$. On the other hand,  to get the local gauge invariance of the model we introduce a link gauge field $U(x,x+e_l)$, where $e_l$ is a unit vector along the  $l$ direction, with $l=1,...,d$,  and a link vector potential $A_a^l(x)=A_a(x,x+e_l)$. Therefore, for each pair of $(x,x+e_l)$ nearest-neighbor lattice sites we write:\\
$$\Psi(x)^+\Psi(x+e_l)\rightarrow\Psi^+(x)U_l(x)\Psi(x+e_l), $$
with
$$U_l(x)=U(x,x+e_l)=e^{iA_a^l(x)T_a}. $$
If lattice sites $(x,y)$ are not nearest-neighbor sites, we consider the link path ordered sequence $$\Gamma(x_1,x_n)=\bigcup_{i=1}^{n-1}(x_i,x_{i+1}), $$ and we define $U(x,y)$ as\\
$$U(x,y)=\prod_{(z,l)\in\Gamma(x,y)}U(z,z+e_l)=e^{i\sum_{(z,l)\in\Gamma(x,y)}A_a^l(z)T_a}, $$\\
where the link gauge field is such that\\
\begin{equation}
\label{b}
F^+U(x,y)F=e^{i\theta_a(x)T_a}U(x,y)e^{-i\theta_a(y)T_a}.
\end{equation} \\

\noindent Now, we introduce the link group generators $E_a^l(x)$ as follows\\
\begin{equation}
F=\prod_{x\in\Lambda}\prod_{l=1}^de^{i\theta_a(x)(\Psi^+(x)T_a\Psi(x)+E_a^l(x))}=\prod_{x\in\Lambda}e^{i\theta_a(x)G_a(x)},
\end{equation}\\
where\\
\begin{equation}
    [G_a(x),G_b(x')]=if_{abc}G_c(x)\delta_{x,x'}.
\end{equation}\\

\noindent Therefore, if $[G_a(x),H]=0$ then the Hamiltonian $H$ exhibits the local gauge symmetry.

As a special case, if $F$ is an infinitesimal transformation, by means of Eqs.~(\ref{a})-(\ref{b}) we get the commutation rules\\
\begin{equation}
\label{com}[G_a,\Psi(x)]=-T_a\Psi(x),
\end{equation}
$$[G_a,\Psi^+(x)]=\Psi^+(x)T_a,$$
$$[G_a,U_l(x)]=-[T_a,U_l(x)],$$\\
with\\
$$G_a=\sum_{x\in\Lambda}G_a(x),$$\\
and for each pair of lattice sites $(x,y)$\\
\begin{equation}
    \label{c}
[G_a(x),U(x,y)]=T_aU(x,y),
\end{equation}
$$[G_a(y),U(x,y)]=-U(x,y)T_a.$$\\

\section{Hubbard model with SOI}
After having considered the main results for the lattice gauge theory in the preceding section, let us now discuss how they can be applied to the Hubbard model in the presence of SOIs.

It is well-known that the Hubbard model describes the electronic motion in the crystals within the tight-binding approximation.~\cite{Hubbard} If one introduces the spin operator $\Psi(x)=\left(\begin{array}{clr}
c_\uparrow(x)\\
c_\downarrow(x)
\end{array}\right)$, where $c_\sigma$ is the annihilation operator at site $x\in \Lambda\subset\mathbb{Z}^d$ for an electron with spin $\sigma$, then the Hubbard Hamiltonian may be written as\\
\begin{equation}
\label{e1}
    H=\sum_{x,y\in\Lambda}t(x-y)\Psi^+(x)\Psi(y)+U\sum_{x\in\Lambda}n_\uparrow(x)n_\downarrow(x),
\end{equation}\\
where $n_\sigma(x)=c_\sigma^+(x)c_\sigma(x)$ is the particle number operator with spin $\sigma$. This Hamiltonian is rotationally invariant in the spin space, i. e. $[H,S^a]=0$, where $S^a=\sum_{x\in\Lambda}S^a(x)$ with\\
$$S^+(x)=c_\uparrow^+(x)c_\downarrow(x) \ \ S^-(x)=c_\downarrow^+(x)c_\uparrow(x),$$
\begin{equation}
\label{com1}
    S^z(x)=\frac{1}{2}(n_\uparrow(x)-n_\downarrow(x)).
\end{equation}

The operators defined by Eq.~(\ref{com1}) are the SU(2) group generators, and their commutation rules are:
\begin{equation}
    [S^+,S^-]=2iS^z\ \ [S^z,S^\pm]=\pm S^\pm.
\end{equation}

According to Mermin-Wagner theorem, a spontaneous magnetic order is absent in the Hubbard model.~\cite{Gosh} This result has been obtained adding to this Hamiltonian of SU(2) symmetry-breaking term, and applying the Bogoliubov inequality.
We notice that the SU(2) symmetry exhibited by the Hubbard model is a global one. To get a SU(2) local gauge symmetry for this model, we may implement the lattice gauge theory previously introduced. In this way the Hubbard Hamiltonian can be written as\\
$$
    H=\sum_{x,y\in\Lambda}t(x-y)\Psi^+(x)U(x,y)\Psi(y)+U\sum_{x\in\Lambda}n_\uparrow(x)n_\downarrow(x)+
    $$
\begin{equation}
\label{e2}
+H_{gauge},
\end{equation}
where $U(x,y)$ is a lattice gauge field and $H_{gauge}$ is the free gauge field Hamiltonian.

Introducing the following field operators\\ $$J^a=\sum_{x\in\Lambda}J^a(x)=\sum_{x\in\Lambda}(\sum_lE^{la}(x)+S^a(x)),$$ i. e. the transformation generators of the local SU(2) group, the Hamiltonian Eq.~(\ref{e2}) is invariant under the considered group, that is $[J^a(x),H]=0$. This result can be easily verified using  Eqs.~(\ref{com})-(\ref{c}), that for SU(2) group are
\begin{equation}
    [J^a,U(x,y)]=-[\frac{\sigma^a}{2},U(x,y)],
\end{equation}
$$[J^a(x),U(x,y)]=\frac{\sigma^a}{2}U(x,y),$$
\begin{equation}
    [J^a(y),U(x,y)]=-U(x,y)\frac{\sigma^a}{2},
\end{equation}
and
\begin{equation}
    [J^a,\Psi(x)]=-\frac{\sigma^a}{2}\Psi(x)\ \ \ \ [J^a,\Psi^+(x)]=\Psi^+(x)\frac{\sigma^a}{2},
\end{equation}
where $\sigma^i$ are the Pauli matrices.

It is well-know that SOIs in condensed matter physics can be described in terms of a non-Abelian gauge theory~\cite{gauge} by means of a non-Abelian potential $A_\mu(x)=\frac{\sigma^a}{2}A_\mu^a(x)$ ($a=1,2,3$) where
$$A_0^a(x)=-\frac{e\hslash}{mc}B^a(x)\ \ A^{ia}(x)=\frac{e\hslash}{mc^2}\epsilon^{iaj}E_j(x). $$
Here, $B^i(x)$ are the components of the external magnetic field and $E^i(x)$ is the electric field produced, for example, by nuclei in molecules or solids, and $a,i,j=1,2,3$.

For 2D models with SOIs interaction of Rashba and/or Dresselhaus type, the time component of SU(2) potential is
$$A^{a0}(x)=\frac{2g_L}{e}\mu_BB^a(x), $$
with an appropriate Land\'e factor $g_L$, and the spatial components are defined as
$$\vec{A}^3(x)=0, $$
\begin{equation}
\label{gauge1}
    \vec{A}^1(x)=\frac{2mc}{e\hslash}\left(\begin{array}{clr}
-\beta\\ \alpha\\ \ 0
\end{array}\right), \ \ \vec{A}^2(x)=\frac{2mc}{e\hslash}\left(\begin{array}{clr}
-\alpha \\ \beta \\ 0
\end{array}\right).
\end{equation}
For instance, in the continuum the substitution
$$\frac{\hat{P}^2}{2m}\rightarrow\frac{1}{2m}(\hat{\vec{P}}-\frac{e}{c}\vec{A}^a\frac{\sigma^a}{2})^2+eA^{a0}\frac{\sigma^a}{2}, $$
allows to obtain the well-known Hamiltonian with SOI Rashba and Dresselhaus couplings in the presence of a magnetic field:
\begin{equation}
\label{hamiltonian}
H=\frac{\hat{P}^2}{2m}+H_R+H_D+g_L\frac{e\hslash}{2mc}\vec{\sigma}\cdotp\vec{B}(x).
\end{equation}
Here,
$$H_R=\frac{\alpha}{\hslash}(P_x\sigma^2-P_y\sigma^1)$$
and
$$H_D=\frac{\beta}{\hslash}(P_x\sigma^1-P_y\sigma^2).$$
are the Rashba and Dresselhaus terms, respectively.

Now, let us introduce SOI in Eq.~(\ref{e1}) by using the SU(2) lattice gauge theory previously outlined.
To this end, we define for our case the link gauge fields as\\
$$U_1(x)=e^{-i\frac{e}{2\hslash c}\sigma^aA^{1a}(x)}=e^{ig(\beta\sigma^1+\alpha\sigma^2)}=$$
$$=\lambda_1 \mathds{1}+i\lambda_2(\cos\phi\sigma^1+i\sin\phi\sigma^2)$$
$$=\left(\begin{array}{clr}
\lambda_1&\lambda_2e^{-i(\phi-\frac{\pi}{2})}\\-\lambda_2e^{i(\phi-\frac{\pi}{2})}&\lambda_1
\end{array}\right), $$
$$U_2(x)=e^{-i\frac{e}{2\hslash c}\sigma^aA^{2a}(x)}=e^{-ig(\alpha\sigma^1+\beta\sigma^2)}=$$
$$=\lambda_1 \mathds{1}-i\lambda_2(\sin\phi\sigma^1+i\cos\phi\sigma^2)=$$
\begin{equation}
\label{e3}
    =\left(\begin{array}{clr}
\lambda_1&-\lambda_2e^{i\phi}\\\lambda_2e^{-i\phi}&\lambda_1
\end{array}\right),
\end{equation}
$$U_3(x)=\mathds{1}, $$
with $g=\frac{m}{\hslash^2}$ $$\lambda_1=\cos(g\sqrt{\alpha^2+\beta^2}), \ \ \lambda_2=\sin(g\sqrt{\alpha^2+\beta^2}), $$\\ and
$\tan\phi=\frac{\alpha}{\beta}$.

\noindent We notice that the link gauge fields verify the following commutation rules:
$$[A^1(x),A^2(x')]=2i(\alpha^2-\beta^2)\sigma^3, $$
\begin{equation}
\label{commutatore}
[U_1(x),U_2(x')]=2i(\lambda_2)^2(\cos^2\phi-\sin^2\phi)\sigma^3.
\end{equation}
It is worth stressing that when the coupling constants $\alpha$ and $\beta$ are equal, the symmetry group becomes an abelian $U(1)$ group. Furthermore, within  the definition reported in Eq.~(\ref{gauge1}), the gauge fields $U_l(x)$ are independent on the choice of the lattice site. This means that, if the pair of lattice sites $(x,y)$ in the hopping term of the Hamiltonian Eq.~(\ref{e2}) are not nearest-neighbor sites, then $U(x,y)$ depends on sites $x$ and $y$ only. But this is not the general case and $U(x,y)$ may depend on the link sequence connecting the $x$ site to the $y$ site. Therefore, if $\Gamma=(x,x_1,...,x_{n-1},y)$ is a path on the lattice, then $U_\Gamma(x,y)=U_{i_1}(x)U_{i_2}(x_1)...U_{i_n}(x_{n-1})$ where $i_k$ is the direction on the lattice connecting $x_{k-1}$ and $x_k$ of the path $\Gamma$. We point out that it is possible to obtain the general form of $U(x,y)$ by imposing the time reversal symmetry of the Hamiltonian in Eq.~(\ref{e2}). The time reversal operator is given by \\
$$T=\prod_{x\in\Lambda}e^{-i\pi S^y(x)}K, $$\\where $K$ is the complex conjugation. By imposing $T^+HT=H$ one obtains
\begin{equation}
\label{Umatrix}
U(x,y)=\left(\begin{array}{clr}
g(x,y)&j(x,y)\\-j^*(x,y)&g^*(x,y)
\end{array}\right).
\end{equation}
Here, the functions $g(x,y)$ and $j(x,y)$ are connected to the electronic jump from $x$ to $y$ with spin conservation and spin flip, respectively. These functions depend obviously on the electronic path followed in the jump.\\
Assuming the local SU(2) gauge symmetry on the lattice, the interaction is contained in the path ordered of the gauge field on the lattice $U(x,y)$, so that the group SU(2) algebra can be used when we apply the Bogoliubov inequality to get the upper bound on the order parameter.

\section{Application and results}
It is important to observe that the spin-orbit coupling breaks explicitly the SU(2) spin symmetry, and therefore we cannot say anything on the spontaneous magnetization. Nevertheless, we will show that within our approach, if Rashba and Dresselhaus coupling constants are equal, then the Hubbard Hamiltonian in Eq.~(\ref{e2}) exhibits a $U(1)$ rotational symmetry in spin space and the SU(2) can be restored by a gauge transformation in the spin space. This conclusion will be used to prove that if hopping matrix $t(x-y)\backsim O(\frac{1}{|x-y|^2})$ then the magnetic ordering is absent in $d=2$, for any  finite temperature, in agreement with Mermin-Wagner theorem. In $d=1$, it is always possible to restore the SU(2) symmetry in spin space by a gauge transformation, and therefore the magnetic order is equally absent. On the other hand, the $\eta$ pairing superconductivity is vanishing for any value assumed by $\alpha$ and $\beta$, because the U(1) symmetry in pseudospin space is not broken at all. To better clarify these points we consider separately the study of the magnetic order in the $d=2$ and the $d=1$ cases, and then the $\eta$ pairing long-range.

To rigorously look at the absence, for finite temperatures, of the spontaneous long-range orders mentioned above, one should first remove the degeneracy of the model Hamiltonian and then study the expectation values involved. The degeneracy is usually removed by introducing a symmetry-breaking term to the Hamiltonian under study. Then, the main steps to follow are: a) the identification of the order parameter involved in the phase transition; b) the addition to the Hamiltonian of a symmetry-breaking term relevant for the order under study; c) the adequate choice of the operators in the Bogoliubov inequality to single-out the order parameter; d) the search for non trivial upper bounds; e) the proof that, in the thermodynamic limit and in the low-dimensional cases, the order parameter vanishes, at any non-zero temperature. The procedure now outlined will be applied in the next subsections to rule out the possibility of magnetic long-range order and then to the superconducting case.

\subsection{Magnetic order in two dimensional lattice with $\alpha=\pm\beta$}
\label{par:2} Here, we show that in the special case when Rashba and Dresselhaus SOI become equal in intensity, the magnetic order is excluded in two-dimensional lattices. When Rashba and Dresselhaus coupling constants are such that  $\alpha=\pm\beta$, the link gauge fields, Eqs.~(\ref{e3}), are:
$$U_1(x)=U_1=\lambda_1\mathds{1}+i\frac{\lambda_2}{\sqrt{2}}(\pm\sigma^1+\sigma^2), $$
\begin{equation}
    \label{link1}
U_2(x)=U_2=\lambda_1\mathds{1}-i\frac{\lambda_2}{\sqrt{2}}(\sigma^1\pm\sigma^2).
\end{equation}

\noindent They are commuting operators $[U_1,U_2]=0$, as it can be deduced looking at Eq.~(\ref{commutatore}). Thus, the symmetry group is the $U(1)$ abelian group of rotations around a given direction. This statement implies that the ordering of the link gauge fields is not important and we can write the hopping term in the Hamiltonian in Eq.~(\ref{e1}) as follows \\
$$\sum_{x,y\in\Lambda}t(x-y)\Psi^+(x)U(x,y)\Psi(y)=$$ $$=\sum_{x,y\in\Lambda}t(x-y)\Psi^+(x)U_1(x)U_1(x+e_1)*$$
$$*...*U_1(x+le_1)U_2(x+le_1)U_2(x+le_1+e_2)...U_1(x+le_1+me_2)*$$
\begin{equation}
*\Psi(x+le_1+me_2),
\end{equation}\\
where $y=x+le_1+me_2$, $e_1$ and $e_2$ being  the unit vectors on the $\hat{x}_1$ and $\hat{x}_2$ axis of the two dimensional lattice, respectively and l and m are integers. Moreover, we can also write
\begin{equation}
\label{link2}
    U(x,y)=U_1^{y_1-x_1}U_2^{y_2-x_2}=e^{ig'(l-m)\vec{n}_\pm\cdotp\vec{\sigma}},
\end{equation}\\
with $g'=g\sqrt{\alpha^2+\beta^2}$ and $\vec{n}_\pm=\frac{1}{\sqrt{2}}(1,\pm 1,0)$, where the sign + (-) holds if $\alpha=\beta$ ($\alpha=-\beta$). It is easy to verify that if $\alpha=\beta$ then the Hamiltonian in Eq.~(\ref{e1}) is rotationally invariant around the $\vec{n}_+=\frac{1}{\sqrt{2}}(1,1,0)$ axis, that is $[\vec{n}\cdotp\vec{S},H]=0$ where $\vec{S}$ is the total spin. This is also easily observable in the Hamiltonian (\ref{hamiltonian}), whereas if $\alpha=-\beta$ the symmetry is around the $\vec{n}_-=\frac{1}{\sqrt{2}}(1,-1,0)$ axis.
After straightforward algebra, the gauge interaction can be written as
\begin{equation}
    U(x,y)=e^{-ig'\theta_\mp(x)\vec{n}_\pm\cdotp\vec{\sigma}}e^{ig'\theta_\mp(y)\vec{n}_\pm\cdotp\vec{\sigma}},
\end{equation}
where $\theta_\pm(x)=\sqrt{2}\vec{n}_\pm\cdotp\vec{x}$.

\noindent Defining a gauge transformation $F$ as
$$F=\prod_{x\in\Lambda}e^{ig'\theta_\mp(x)\vec{n}_\pm\cdotp\vec{S}(x)}, $$
\noindent we find that the Hubbard Hamiltonian with spin orbit coupling is globally SU(2) invariant in the spin space:
\begin{equation}
    F^+\Psi^+(x)U(x,y)\Psi(y)F=\Psi^+(x)\Psi(y).
\end{equation}
If we consider the average value of the spin projection along $\hat{x}_3$ axis, we can write
$$<S^3(x)>=tr(e^{-\beta H}S^3(x))=tr(F^+e^{-\beta H}FF^+S^3(x)F)=$$ $$=tr(e^{-\beta F^+HF}\vec{\rho}\cdotp\vec{S}(x)), $$
where $\rho_i=R_{i3}(e_3)_i$ and $R$ is a rotation matrix of $\theta_\mp(x)$ around $\vec{n}_\pm$ axis.
The Hamiltonian is now SU(2) invariant, and then by the global gauge transformation $O=e^{-ig'\theta_\mp(x)\vec{n}_\pm\cdotp\vec{S}}$ we get
$$<S^3(x)>=<S^3(x)>', $$
where $<\ \ >'$ is the average value of the $S^3(x)$ in the transformed Hamiltonian.

Since for the Hubbard model without spin orbit coupling the spontaneous magnetization is absent,\cite{Gosh} introducing the magnetic order parameter as
$$m_h(Q,\Lambda)=\frac{1}{|\Lambda|}\sum_{x\in\Lambda}e^{iQ\cdotp x}<S^3(x)>=\frac{1}{|\Lambda|}<S^3(Q)>', $$
we trivially deduce
$$\lim_{h\rightarrow 0}\lim_{|\Lambda|\rightarrow\infty} m_h(Q,|\Lambda|)=0. $$

\subsection{Magnetic order in one dimensional lattice}
\label{par:1} We consider the Hubbard model on a chain of lattice sites  with a constant lattice $a$ assumed for simplicity equal to 1. Therefore, the crystal lattice is $\Lambda\subset\mathbb{Z}$, and the Hubbard Hamiltonian is
$$H=\sum_{l=1}^{|\Lambda|}\sum_{j=1}^{|\Lambda|-l}t(j)\Psi^+(l)\Psi(l+j)+U\sum_{l=1}^{|\Lambda|}n(_\uparrow)n(_\downarrow l). $$
In one spatial dimension, we have one link direction, so that the lattice may be seen as a sequence of links. Also, it can be considered as  U(1) subgroup of SU(2) where we have one gauge field associated with the link ($l,l+1$), that is, the group is abelian and its elements are:
$$U_1(x)=e^{-i\frac{e}{\hslash c}\sigma^aA^{1a}}=\lambda_1\mathds{1}+i\lambda_2(\cos\phi\sigma^1+\sin\phi\sigma^2). $$
To introduce the spin-orbit coupling we perform the replacement
$$\Psi^+(l)\Psi(l+j)\rightarrow\Psi^+(l)U_1(l)U_1(l+1)...U_1(l+j-1)\Psi(l+j)=$$
$$=\Psi^+(l)U(l,l+j)\Psi(l+j). $$\\
Since the gauge fields are independent on the site, we write
\begin{equation}
\label{link}
    U(l,l+j)=e^{ig'j\vec{n}\cdotp\vec{\sigma}},
\end{equation}
with $g'=g\sqrt{\alpha^2+\beta^2}$ and $\vec{n}=(\cos\phi , \sin\phi,0)$.

Summarizing, the Hubbard Hamiltonian is rotationally invariant around the vector $\vec{n}$.  Indeed if $\vec{S}=\sum_{l=1}^{|\Lambda|}\vec{S}(l)$ denotes the total spin operator, then we have $[\vec{S}\cdotp\vec{n},H]=0$. Furthermore, Eq.~(\ref{link}) is very similar to Eq.~(\ref{link2}), so that we may define the unitary operator $F=\prod_{l=1}^{|\Lambda|}e^{ig'l\vec{n}\cdotp\vec{S}(l)}$. Therefore, we can write
$$F^+\Psi^+(l)U(l,l+j)\Psi(l+j)F=\Psi^+(l)\Psi(l+j), $$
that is, the transformed Hamiltonian is the Hubbard model without the spin-obit interaction. Therefore, using the argument previously outlined for the two-dimensional lattice, we can conclude that the spontaneous magnetization is absent for any $\alpha$ and $\beta$ values, also in the 1D case.

\subsection{$\eta$ pairing long-range order in one and two dimensional lattices}

The $\eta$ pairing long-range order is connected to the spontaneous symmetry breaking of the $U(1)$ invariance around the third axis in the pseudospin space of the Hamiltonian.~\cite{Yang} The Hubbard Hamiltonian exhibits this symmetry and the introduction of the spin orbit interaction does not modify this property. If we define the pseudospin operators as
$$\eta^-(x)=c_\uparrow (x)c_\downarrow(x)\ \ \eta^+(x)=c_\downarrow^+(x)c_\uparrow^+(x), $$
$$\eta^3(x)=\frac{1}{2}(n_\uparrow(x)+n_\downarrow(x)-1), $$
then $[\eta^3,H]=0$. Here,\\ $$\eta^3=\sum_{x\in\Lambda}\frac{1}{2}(n_\uparrow(x)+n_\downarrow(x)-1). $$
From the commutators
$$[\eta^3,U_1]=[\eta^3,U_2]=0, $$
\begin{equation}
\label{commeta}
    [\eta^3,\Psi(x)]=-\Psi(x) \ \ [\eta^3,\Psi^+(x)]=\Psi^+(x),
\end{equation}\\
it is easy to prove that\\ $$[\eta^3,\Psi^+(x)U(x,y)\Psi(y)]=0. $$
The absence of long-range $\eta$ pairing in the Hubbard model without spin-orbit coupling has been widely studied.~\cite{etap} Here, following the same approach, we extend this result to the Hubbard Hamiltonian with SOI. To this end, we introduce the symmetry breaking external field $\lambda$ in the Hamiltonian as follows
$$H\rightarrow H-\lambda(\eta^+(Q)+\eta^-(-Q)), $$
and we define the $\eta$ pairing order parameter as
$$\Delta(Q)=\lim_{|\Lambda|\rightarrow\infty}\frac{1}{|\Lambda|}\sum_{x\in\Lambda}
e^{-iQ\cdotp x}<\eta^+(x)>=$$ $$=\lim_{|\Lambda|\rightarrow\infty}\frac{<\eta^+(Q)>}{|\Lambda|}. $$
To show that long-range $\eta$ pairing is absent in this model we have to show that $$\lim_{\lambda\rightarrow 0}\Delta_\lambda=0. $$ So, in the Bogoliubov inequality
\begin{equation}
    \label{Bog1}
|<[A,B]>|^2\leqslant \frac{1}{2k_BT}<[A,A^+]_+><[B^+,[B,H]]>,
\end{equation}
we define the operators $A$ and $B$ as
$$A(q)=\eta^+(q+Q)\ \ B(q)=\eta^3(-q). $$
Then, from the commutator
$$[B,\Psi^+(x)U(x,y)\Psi(y)]=(e^{iq\cdotp x}-e^{iq\cdotp y})\Psi^+(x)U(x,y)\Psi(y), $$
we get the average value of the double commutator as
$$<[B^+,[B,H]]>=$$ $$=2\sum_{x,y\in\Lambda}t(x-y)(1-\cos(q\cdotp(x-y)))*$$ $$*<\Psi^+(x)U(x,y)\Psi(y)>-\lambda(<\eta^+(Q)>+<\eta^-(-Q)>). $$
Now, if one defines the scalar product by $(A,B)$=$<A^+B>$, then by Schwartz inequality one gets
$$<\Psi^+(x)U(x,y)\Psi(y)>\leqslant 2, $$
so that
\begin{equation}
    \label{dis2}
|<[B^+,[B,H]]>|\leqslant|\Lambda|(\frac{2q^2}{\rho}+\lambda\Delta(Q,\Lambda)).
\end{equation}\\
Since $<[A,B]>=|\Lambda|\Delta(Q,\Lambda) $ and
$$\sum_{q}<[A,A^+]_+>\leqslant|\Lambda|^2, $$ by using the Bogoliubov inequality Eq.~(\ref{Bog1}) the proof is accomplished. Indeed, we get $$|\Delta(Q)|^2\int\frac{d^dq}{(2\pi)^d}\frac{1}{q^2+\lambda\frac{\rho}{2}\Delta(Q)}\leqslant\frac{2}{\rho k_BT}, $$
Thus, by solving the integral we find out that, when $\lambda \rightarrow 0$, $\Delta(Q) \rightarrow 0$
for $d$=1, 2, at finite temperature. This implies that the Hubbard model with SOI does not exhibit the $\eta$ pairing long-range order.

\section{Conclusions}
We presented an extension of the Mermin-Wagner theorem for the Hubbard model in the presence of SOI, and showed that spontaneous magnetic order is ruled out in two dimensions, at finite temperature, if the Rashba ($\alpha$) and the Dresselhaus ($\beta$) spin-orbit interactions are such that $\alpha=\pm \beta$. On the contrary, in one-dimension the magnetic order can be excluded, regardless of the values assumed by the spin-orbit coupling constants.
We notice that, when $Q$=0 in $m(Q)$ the ferromagnetic order is forbidden, while choosing $Q$ in such a way that $\exp(i{Q R}_i)=\pm 1$ when ${R}_i$ connects sites in the same sublattice and different sublattices, respectively, we argue that also the antiferromagnetic order is forbidden.
We also proved the absence of long-range $\eta$ pairing, at finite temperatures, in one- and two-dimensions, independently on  $\alpha$ and $\beta$ interaction parameters. As stated for the magnetic order, looking at the $\eta$-pairing order parameter $\Delta(Q)$,  we may infer that for $Q=0$ the s-wave pairing can be excluded, for $Q=\pm {\pi}$ we exclude the $\eta$-pairing, and finally for $ Q\neq \{0, \pm {\pi}\}$ we rule out the existence of generalized $\eta$-pairing order with momentum $Q$.

For copleteness, we note that the Mermin-Wagner theorem follows from the fact that in low-dimensional cases, a diverging number of infinitesimally low-lying excitations is created at any finite temperature, and thus the assumption of a non vanishing order parameter is not self-consistent. This consideration, as well as the rigorous proof, does not apply at T=0, implying that the ground-state may be ordered. For instance, two-dimensional ferro(anti)magnetism is possible at zero temperature: quantum fluctuations oppose but do not prevent the appearance of a two-dimensional magnetically ordered phase. In contrast, for one-dimensional systems quantum fluctuations become so strong that they usually prevent even ground state ordering. Indeed, it is known that the ground-state of the one-dimensional Hubbard model is a non magnetic singlet at any band filling and for any value of Coulomb interaction U. More generally, if the energy spectrum has a gap, it can be shown that the model under investigation does not exhibit long range order, and interestingly, this energy gap plays the role of the temperature in conventional Bogoliubov inequality.~\cite{auerbach, su97,noce05,noce06}

\section*{Acknowledgements}
We would like to express our gratitude to  Mario Cuoco and  Alfonso Romano for helpful comments and valuable discussions.


\begin{thebibliography}{999}
\bibitem{Dys} F. J. Dyson, E. H. Lieb, and B. Simon, Phys. Rev. Lett. {\bf 37}, 120 (1976);  J. Stat. Phys.  {\bf 18},  335  (1978).

\bibitem {bog} N. N. Bogoliubov, Physica {\bf 26}, S1 (1960); Phys. Abh. Sowjetunion {\bf 6}, 1, 113, 229 (1962).

\bibitem{Nolt}A. Gelfert and W. Nolting, J. of Phys.: Condens. Matter {\bf 13} R505 (2001).

\bibitem{MW}  N. D. Mermin and H. Wagner, Phys. Rev. Lett. {\bf 17}, 1133 (1966).

\bibitem{RKKY} M. A. Rudemann and C. Kittel, Phys. Rev. {\bf 96}, 99 (1954); T. Kasuya, Prog. Theor. Phys. {\bf 16}, 45 (1956); K. Yoshida,
Phys. Rev. {\bf 106}, 893 (1957).

\bibitem{Bruno} P. Bruno, Phys. Rev. Lett. {\bf 87}, 137203 (2001).

\bibitem{Fes} D. Loss, F. L. Pedrocchi, and A. J. Leggett, Phys. Rev. Lett. {\bf 107}, 107201 (2011).

\bibitem{Ras} E. I. Rashba, Phys. Rev. B {\bf 68}, 241315(R) (2003).

\bibitem{Dress} G. Dresselhaus, Phys. Rev. {\bf 100}, 580 (1955).

\bibitem{Mai} L. Meier, G. Salis, I. Shorubalko, E. Gini, S. Sch¨on, and K. Ensslin, Nature Physics {\bf 3}, 650 (2007).

\bibitem{El} S. Elitzur, Phys. Rev. D {\bf 12}, 3978 (1975).

\bibitem{Higgs} P. W. Anderson, Phys. Rev. {\bf 130}, 439 (1963); P. W. Higgs, Phys. Lett. {\bf 12}, 132 (1964); Phys. Rev. Lett. {\bf 13}, 508 (1964); Phys. Rev. {\bf 145}, 1156 (1966) F. Englert and R. Brout, Phys. Rev. Lett. {\bf 13}, 321 (1964); G. S. Guralnik, C. R. Hagen, and T. W. B. Kibble, ibid. {\bf 13}, 585 (1964); T. W. B. Kibble, Phys. Rev. {\bf 155}, 1554 (1967).

\bibitem{yanase04} Y. Yanase, T. Jujo, T. Nomura, H. Ikeda, T. Hotta and K. Yamada, Phys. Rep. {\bf 387}, 1 (2004); T. Yokoyama, S. Onari, Y. Tanaka, J. Phys. Soc. Jpn. {\bf 77}, 064711 (2008); Y. Yanase and M. Sigrist, J. Phys. Soc. Jpn. {\bf 77}, 124711 (2008).

\bibitem{yokoyama07} T. Yokoyama, S. Onari and Y. Tanaka, Phys. Rev. B {\bf 75}, 172511 (2007).

\bibitem{yada09} K. Yada, S. Onari and Y. Tanaka, J. I. Inoue, Phys. Rev. B {\bf 80}, 140509(R) (2009).

\bibitem{alexandradinata10} A. Alexandradinata and J. E. Hirsch Phys. Rev. B {\bf 82}, 195131 (2010).

\bibitem{KS} J. Kogut and L. Susskind, Phys. Rev. D {\bf 11}, 395 (1975); R. Balian, J. M. Drouffe, and C. Itzykson, Phys. Rev. D {\bf 10}, 3376 (1974); R. Balian, J. M. Drouffe, and C. Itzykson Phys. Rev. D {\bf 11}, 2098 (1975).

\bibitem{Hubbard} For a review see: M. Rasetti ed., {\it The Hubbard model: recent results} (World Scientific, Singapore, 1999).

\bibitem{gauge} Y. Aharonov and A. Casher, Phys. Rev. Lett. {\bf 53}, 319 (1984); J. Anandan, Phys. Lett. A {\bf 138}, 347 (1989); H. Mathur and  A. D. Stone, Phys. Rev. Lett. {\bf 68}, 2964 (1992); Y. Oreg and O. Entin-Wohlman, Phys. Rev. B {\bf 46}, 2393 (1992); Pei-Qing Jin, You-Quan Li, and Fu-Chun Zhang, J. Phys. A: Math. Gen. {\bf 39}, 7115 (2006).

\bibitem{Gosh} D. K. Ghosh, Phys. Rev. Lett. {\bf 27}, 1584 (1971).

\bibitem{etap} G. Su, A. Schadschneider, and J. Zittartz, Phys. Lett. A {\bf 230}, 99 (1997).

\bibitem{Yang} C. N. Yang and S. C. Zhang, Mod. Phys. Lett {\bf 34}, 759 (1990).

\bibitem{auerbach}A. Auerbach, {\it Interacting electrons and quantum magnetism} (Springer Verlag, Berlin 1994).

\bibitem{su97}  G. Su, A. Schadschneider, and J. Zittartz, Phys. Lett. A {\bf 230}, 99 (1997).

\bibitem{noce05} C. Noce, Phys. Rev. B {\bf 71}, 092506, (2005).

\bibitem{noce06} C. Noce, Phys. Rep. {\bf 431}, 173 (2006).


\end{thebibliography}
\end{document}